\documentclass[aps,prl,twocolumn,groupedaddress,showpacs]{revtex4}
\usepackage{graphicx}

\begin{document}
\title{Low velocity quantum reflection of Bose-Einstein condensates}
\author{T.~A. Pasquini, M. Saba, G. Jo, Y. Shin, W. Ketterle,  D.~E. Pritchard}
\homepage{http://cua.mit.edu/ketterle_group/}

\affiliation{Department of Physics, MIT-Harvard Center for
Ultracold Atoms, and Research Laboratory of Electronics,
Massachusetts Institute of Technology, Cambridge, Massachusetts,
02139}
\author{T.~A. Savas}

\affiliation{NanoStructures Laboratory and Research Laboratory of
Electronics, Massachusetts Institute of Technology, Cambridge,
Massachusetts, 02139}

\author{N. Mulders}

\affiliation{Department of Physics, University of Delaware, Newark,
DE, 19716}
\date{\today}

\begin{abstract}

We studied quantum reflection of Bose-Einstein condensates at normal
incidence on a square array of silicon pillars. For incident
velocities of 2.5-26~mm/s observations agreed with theoretical
predictions that the Casimir-Polder potential of a reduced density
surface would reflect slow atoms with much higher probability. At
low velocities (0.5-2.5~mm/s), we observed that the reflection
probability saturated around 60\% rather than increasing towards
unity. We present a simple model which explains this reduced
reflectivity as resulting from the combined effects of the
Casimir-Polder plus mean field potential and predicts the observed
saturation. Furthermore, at low incident velocities, the reflected
condensates show collective excitations.

\end{abstract}


\maketitle

Quantum fluctuations of the electromagnetic field exert forces on
objects and are responsible for the attractive interactions between
two neutral objects, e.g. an atom and a surface~\cite{CAS48}.  Such
interactions are typically weak and decay rapidly with increasing
separation. Still, they are important for nanoscale
devices~\cite{BOR01, CHEN01} and when ultracold atoms are trapped
close to a surface~\cite{HAR03, MCG04, LIN04}. One spectacular
consequence of the Casimir-Polder potential is the prediction of
total quantum reflection of very slow atoms from neutral surfaces:
atoms incident on a surface at low velocity are accelerated toward
the surface so abruptly that they reflect from the potential instead
of being drawn into the surface~\cite{CLO92, CAR98, MODY01, FRI02}.
If high reflection probabilities could be realized, new atom-optical
devices such as mirrors and cavities would be possible without the
need of magnetic or optical fields. However, in a recent study of
quantum reflection of Bose-Einstein condensates, the reflection
probability was limited to $\sim15\%$ at low velocity~\cite{PAS04}.
A theoretical paper simulating quantum reflection of Bose-Einstein
condensates could not explain the low reflectivity~\cite{SCO05}.

In this work, we investigate the quantum reflection of Bose-Einstein
condensates (BECs) from surfaces for velocities near and below
1~mm/s. To enhance quantum reflection, we use a pillared silicon
surface, in the spirit of previous experiments with
grazing-incidence neon atoms on ridged silicon~\cite{SHI02a, OBE05,
KOU05}. We observe quantum reflection of Bose-Einstein condensates
with probabilities of up to 67$\%$ for velocities of $\sim$1~mm/s,
corresponding to a collision energy of $k_B \times$ 1.5~nK. We
propose a simple model to explain how mean field interactions
interfere with the reflection process and prevent the observation of
higher reflection coefficients with BECs. Further, due to the
greatly enhanced reflection coefficients, we observe collective
excitations of the reflected condensate and incoherent scattering
between incident and reflected clouds.

\begin{figure}
\begin{center}
\includegraphics{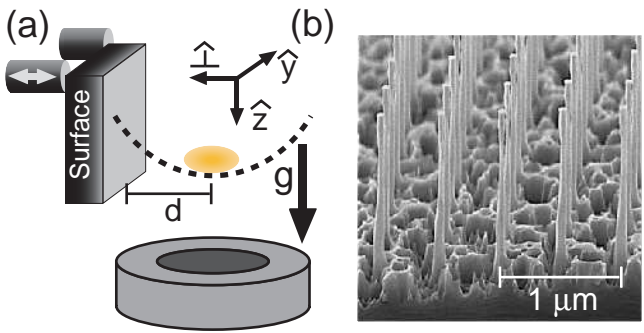}
\caption{Experimental schematic. (a) Atoms were confined in a
gravito-magnetic trap near a pillared Si surface. Atoms were
accelerated towards the surface by displacing the trapping potential
a distance $d$ (greatly exaggerated) so that it was centered on the
surface. The surface was mounted on a translation stage and could be
removed at any point for imaging. (b) Scanning electron micrograph
of the pillared Si surface used in this
experiment.\label{f:schematic}}
\end{center}
\end{figure}

Bose-Einstein condensates of $^{23}$Na atoms were prepared and
transferred into a loosely confining gravito-magnetic trap,
comprising a single coil and three external bias fields, as
described in Ref.~\cite{LEA03a}. For typical loading parameters,
condensates with $N\approx1\times10^6$ atoms were confined
$\sim$1~cm above the coil in a harmonic trap characterized by
angular frequencies $(\omega_\bot, \omega_y, \omega_z)=2\pi \times
(4.2, 5.0, 8.2)$ Hz, where directions $(\bot, y, z)$ are defined in
Figure~\ref{f:schematic}. At this point, $\omega_\bot$ and
$\omega_y$ were adjusted by changing the vertical bias field as
described in Ref.~\cite{LEA03a}. Typical densities in the trap were
$\sim5\times10^{12}$~cm$^{-3}$ and diameters were $\sim150~\mu$m. A
silicon surface attached to a micrometric, motorized linear actuator
was mounted $\sim$1~cm above the single coil. The position of the
surface relative to the center of the coil was adjustable during the
experiment as shown in Figure~\ref{f:schematic}a.

The surface used in this experiment, provided by the MIT
Nanostructures laboratory, was a pillar structure etched into
single-crystal silicon.  The structure was created by  Interference
Lithography (IL) and various subsequent etching steps~\cite{SMI91,
SCH95, SAV96}.  Figure \ref{f:schematic}b shows the final surface as
an array of 1~$\mu$m tall, 50~nm diameter pillars spaced at 500~nm.
Such a surface should provide a Casimir potential approximately 1\%
of the value for a solid Si surface.  The quantum reflection
efficiency depends inversely on the strength of the interaction, and
a dilute surface is expected to exhibit enhanced reflection.

\begin{figure}
\begin{center}
\includegraphics{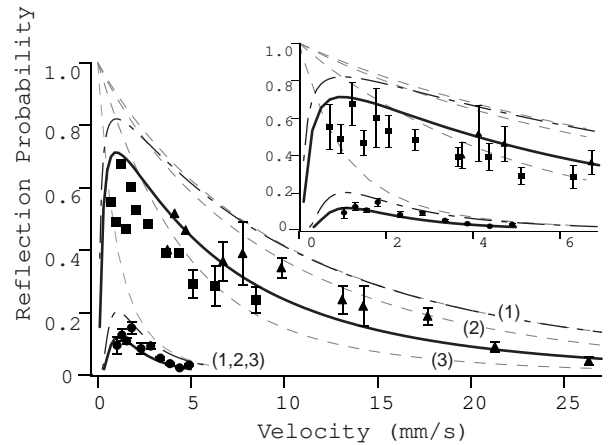}
\caption{Reflection probability vs incident velocity.  Data were
collected in a magnetic trap with trap frequencies $2\pi \times
(2.0, 2.5, 8.2)$~Hz (squares) and $2\pi \times (4.2, 5.0, 8.2)$~Hz
(triangles). For comparison, data from Ref.~\cite{PAS04} for
reflection off a solid silicon surface are shown as circles.
Incident and reflected atom numbers were averaged over several
shots. For clarity error bars for data below 5~mm/s are shown only
on the inset plot, which has a different horizontal axis to
emphasize the low velocity data. Systematic uncertainty in the
velocity due to residual motion is approximately 10\%. Theoretical
curves are described in the text. \label{f:reflectivity}}
\end{center}
\end{figure}

Studying the reflection properties of the surface requires a
controlled collision. After loading the condensate into the trap,
the surface was moved to a desired distance $d$ from the trap
center. By changing the bias field $B_\bot$ appropriately, a dipole
oscillation centered on the surface was induced~\cite{PAS04}. After
waiting $T_\bot/4=2\pi/4\omega_\bot$ the atoms hit the surface with
velocity $v_\bot=d\omega_\bot$. By varying $\omega_\bot$ between
$2\pi\times$2 and $2\pi\times$4~Hz and $d$ over 50~$\mu$m to 1~mm,
velocities in the range of 0.5 to 26~mm/s could be studied. The
reflection probability was calculated as the ratio of the average
reflected atom number to the average incident atom
number~\footnote{Atom numbers were obtained by exposing the atoms to
a light pulse which transferred atoms from the state $|F=1\rangle$
to $|F=2\rangle$. After a wait time of $\sim$2~ms to allow optically
dense regions of the cloud to expand, the distributions were imaged
using the $|F=2\rangle$ to $|F=3\rangle$ cycling transition.
Expansion by photon recoil distorts the distribution, but gives
accurate relative numbers.}.The reflection probability, along with
data previously collected for a solid silicon surface~\cite{PAS04},
are shown in Figure~\ref{f:reflectivity}. The pillared surface shows
higher reflectivity over a wider range of incident velocity, as
expected. The reflection maximum is 67$\%$ for a velocity of
1.2~mm/s and reflection probabilities above 10$\%$ were measured at
velocities up to 20~mm/s. Below $\sim$3~mm/s, the reflection
probability flattens near 55$\%$, qualitatively similar to the
behavior of the solid surface where the reflectivity flattened near
12$\%$ in the same velocity range.

We calculated a theoretical reflection probability of a single atom
from the pillared surface using three numerical simulations. The
surface potentials of the Casimir-Polder form $C_4/r^4$ are obtained
using C$_4^{Si}= 6.2\times10^{-56}$~Jm$^4$ for bulk
silicon~\cite{YAN97} and combining contributions from both the
pillar layer and the bulk substrate. Reflection probabilities were
calculated by numerically solving the Schr\"{o}dinger equation for a
1D potential~\cite{SHI01}. We consider three averaging schemes for
simulating the experiment: (1) we average the density of the
material before calculating the potential, simulating the surface as
a 1~$\mu$m thick overlayer of material with
C$_4=0.01\times$C$_4^{Si}$ added to a semi-infinite slab of material
with C$_4=$C$_4^{Si}$, (2) we calculate the 3D potential from the
pillared structure numerically by integrating over the regions of
space containing material with C$_4=$C$_4^{Si}$~\cite{MIL92} and
then average to obtain a 1D potential, or (3) we assume that atoms
follow a linear trajectory toward the surface (Eikonal
approximation), and calculate the reflection probability from many
points above the surface before averaging the reflection
probability. The resulting reflection probability curves are shown
in Figure~\ref{f:reflectivity} as numbered dashed gray lines.   The
predictions of model (1) and (2) are similar for the pillared
surface.  They show that the reflection probability depends mainly
on the the diluted pillar layer and only weakly on the bulk material
underneath or the arrangement of the pillars. Model (3) should only
be valid for high incident velocity, when the de Broglie wavelength
$\lambda_{dB}$ is significantly smaller than the surface structure.
This is not the case in our experiment where
$\lambda_{dB}\simeq1\mu$m exceeds the spacing of the pillars.

All calculations predict that the reflection probability approaches
unity for low incident velocity. This is in contrast to our
observation that the reflection probability saturates below 3~mm/s
for both the pillared and solid surfaces.  It was suggested that
this saturation is due to low velocity excitations which smear out
the condensate density. Although the reflectivity approaches unity,
some reflected atoms would appear in a diffuse cloud which may fall
below a detection threshold~\cite{SCO05}. However, this could
explain our previous results~\cite{PAS04} only when we assume a
density threshold for detection of $0.25\times n_0\approx10^{12}$~
cm$^{-3}$, where $n_0$ is the central condensate density, which is
twenty times higher than the lowest densities we are able to detect
~\cite{LEA03a}.

There is a finite-size correction to the standard description of
quantum reflection, but it is too small to account for our
observations. For an incident atom cloud of size $d$, the smallest
incident velocity is $h/md$, approximately 0.2 mm/s for our
parameters. We conclude that a single-particle description can not
account for our low-velocity data and now discuss possible effects
due to the condensate's mean field interaction.

\begin{figure}
\begin{center}
\includegraphics{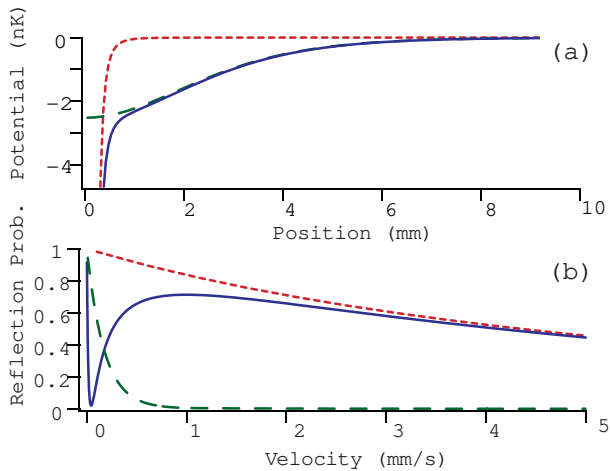}
\caption{Mean field model for quantum reflection of condensates. (a)
The trapped condensate provides a repulsive mean-field energy which
is a constant away from the surface and, within the healing length
$\xi$, drops to zero.  The dashed curve shows this mean-field
potential set to zero at infinity. This potential combined with the
Casimir potential (dotted), creates the composite potential (solid)
which we use to model reflection in the presence of a condensate.
(b) The reflection probabilities from the same the potentials for
low velocities. \label{f:model}}
\end{center}
\end{figure}

The mean field potential is taken to be that of a condensate at rest
with a fully reflecting wall as a boundary condition.  The
condensate's density decays towards zero at the wall over a
characteristic length scale given by the healing length, $\xi$.  The
atoms at the edge of the condensate thereby acquire a velocity given
by $\approx h/m \xi$  which is approximately equal to the speed of
sound $c$. If the healing length is much larger than the relevant
range of the Casimir-Polder potential, approximately 1~$\mu$m as
defined by the so called badlands region~\cite{FRI02}, one would
assume that the mean field potential simply accelerates the atoms.
Atoms leaving the condensate enter the region of quantum reflection
with an incident velocity obtained from $mv^2/2=U=mc^2$. This model
would shift the single-atom quantum reflection curves by the
velocity $v=\sqrt{2}c$ which is $\approx1.5$~mm/s for our
parameters. This shift is too small to explain the low reflectivity
at our lowest velocities. Additionally, the assumption that the
healing length is much larger than the distance at which quantum
reflection happens is not valid for our data.

In order to fully account for interaction effects, we calculate the
quantum reflection probability using a composite potential which
includes both the Casimir-Polder potential and the mean field
potential (Fig~\ref{f:model}a). The results now show a dramatic
reduction of the reflectivity at low velocity, as shown in
Figure~\ref{f:model}b. At high velocities ($>3$~mm/s), quantum
reflection occurs close to the surface where the mean field
potential plays no role. As a result, the predictions of the
composite model are similar to the single atom theory.  As the
velocity is reduced, the point of reflection moves outward, into the
region where the mean-field potential ``softens'' the Casimir-Polder
potential. At very low velocities ($<$0.1~mm/s), when the badlands
region is far from the surface, the predicted reflection resembles
the reflection probability from the tail of the condensate rather
than from the Casimir-Polder potential. We note that the model
continues to predict perfect reflection at zero incident velocity,
however the approach is distinctly non-monotonic; the reflection
probability saturates and decreases precipitously near 1~mm/s before
rising to unity. This model predicts well, without any free
parameter, the velocities below which we have observed saturation of
the reflectivity for both the solid and pillared surface as shown by
the dot-dashed lines in Figure~\ref{f:reflectivity}. Unfortunately,
the data do not extend far enough into the low velocity regime to
confirm the model's prediction of a sharp drop at low velocity or
the ultimate asymptote to unity.

The model does not include the effects of the moving condensate, its
observed collective excitations, or the distortion of the condensate
wavefunction by surface attraction or the loss of atoms to the
surface.

The calculated curves are not in quantitative agreement with the
experimental data; the observed reflection probabilities are lower,
even at high velocity.  One possible explanation is the further
modification of the potential by stray electric fields, caused by
sodium atoms deposited on the surface (adatoms). Recently, the
partial ionization of rubidium adatoms by bulk silicon has been
shown empirically to produce an electric field of several V/cm at
10~$\mu$m from the surface~\cite{MCG04}. This electric field, which
falls off as $1/r^2$, will produce an additional potential,
$V_A(r)=-A/r^4$, which will reduce the reflection probability. To
account for  stray electric fields, we fit the high velocity data
for the pillared (solid) surface using a potential
$V_{tot}=-0.01\times C_4^{Si}/r^4 - A/r^4$ ($V_{tot}=-C_4^{Si}/r^4 -
A/r^4$). We find for the pillared (solid) surface a value of $A$ of
$0.02\times C_4^{Si}$ ($C_4^{Si}$) corresponding to a stray field
$\sim$10~V/cm ($\sim$100~V/cm) at 1~$\mu$m for the pillared (solid)
surface, smaller than the values measured in the rubidium
experiment. If we combine the stronger surface potential with the
mean-field potential we have a phenomenological model which is
consistent with all our data, shown in Figure~\ref{f:reflectivity}
as solid lines. It would be very interesting to test this model by
varying the density over a large range and try to observe the
predicted decrease of the saturation velocity for lower density.
Unfortunately, we couldn't study quantum reflection at lower density
due to rapid decrease of the signal-to-noise ratio.

\begin{figure}
\begin{center}
\includegraphics{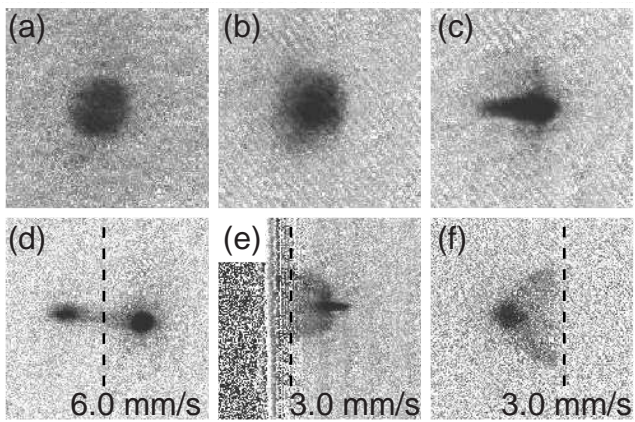}
\caption{Menagerie of reflection effects. (a,b,c) As the incident
velocity is reduced, the reflected condensate becomes increasingly
excited. (d) By removing the surface at the moment of reflection, we
can see both the incident (left) and reflected (right) condensates.
The reflection probability is 30\%. (e) The collision of the
incident and reflected condensates produces a strong s-wave
scattering halo at low velocity, visible here T$_\bot$/4 after
reflection. The surface is still present on the left in this image.
Half of the halo is missing due to surface reflection or absorption.
(f) With the surface removed, the scattered atoms remain in the trap
after an additional half trap period, and appear reversed in
position and velocity. Field of view for images a, b, c is
540~$\mu$m and for d, e, f is 800~$\mu$m; the dashed line is the
position of the surface (moved for imaging) at the moment of
reflection. \label{f:menagerie}}
\end{center}
\end{figure}

The higher reflection efficiency of the pillared surface in excess
of $50\%$ allowed us to study other aspects of quantum reflection of
condensates at low velocity. Because of the finite size of the
condensate, a standing wave is present in the condensate during the
collision time which is inversely proportional to the incident
velocity. As this time becomes comparable to the transverse and
vertical trap periods, vortex rings, solitons, and other excitations
may form, distorting the cloud~\cite{SCO05}. In our experiment, this
velocity is approximately 2mm/s.  At high velocities $(>4$mm/s), we
observe that the reflected condensate appears, apart from diminished
size and number, similar to the incident condensate. A condensed
fraction and a thermal population, both present in the initial cloud
are also present in the reflected cloud, shown in
Figure~\ref{f:menagerie}a. As the incident velocity is reduced, as
in Fig.~\ref{f:menagerie}b and c, the cloud develops a complex
surface mode excitation~\footnote{A movie of reflection at 3~mm/s is
available at http://cua.mit.edu/ketterle\_ group/Animation\_folder
/QRMovie.wmv.}.

Furthermore, we observe elastic s-wave scattering between atoms in
the incident and reflected condensates. S-wave scattering
redistributes atoms in two colliding clouds with initial relative
wavevector $\vec{k}$ evenly onto a sphere of radius
$|\vec{k}|$~\cite{GIB95, CHI00}; an image of the atoms taken after a
hold time T$_\bot$/4 will show the two clouds on opposite sides of
the scattering sphere. In the present experiment, the reflected
front part of the condensate collided with the still incident tail
part (Fig.~\ref{f:menagerie}d). The scattering halo was observed
after sufficient hold time (Fig.~\ref{f:menagerie}~e,f).

We also performed the experiment using an aerogel surface. Aerogels
are electrically insulating, randomly structured, silica foams with
a density of $\sim$2\% of bulk silica~\cite{FRIC92} and should
display reflection properties similar to the pillared surface. We
were unable to observe quantum reflection above our detection
threshold of $\sim$2\%, an effect we attribute to uncontrolled patch
charges which strongly distort the Casimir-Polder potential and
prevent efficient reflection.

In light of the strong increase in reflection probability for the
pillared structure, we want to discuss the ultimate limits of
quantum reflection probability for condensates.  We have found
strong evidence that the presence of a condensate will distort the
potential and prevent efficient reflection at low velocity. Our
simple model predicts improvements for longer healing lengths.
Unfortunately, the corresponding reduction in condensate density
would be a severe limit for atom optical devices based on quantum
reflection. Another way to improve the reflectivity is by further
reducing the density of the surface. Certainly it is possible to
increase the pillar spacing.  However, such widely spaced pillars
will only dominate the potential if their height is simultaneously
increased, which is beyond the limit of current fabrication
techniques. Similarly, narrower pillars may be possible, but not
with the height required to dominate the potential.

This work was supported by NSF, ONR, ARO, DARPA, and NASA. We thank
S. Will for experimental assistance, M. Zwierlein for suggestions
about the experiment and R. Scott and M. Fromhold for useful
discussions.

\end{document}